\documentclass[english,twoside,a4paper,10pt]{article}

\usepackage[latin1]{inputenc}
\usepackage[T1]{fontenc}
\usepackage{amsmath}
\usepackage{amsfonts}
\usepackage{graphicx}
\usepackage{subfigure}
\usepackage{a4wide}
\usepackage{amssymb}
\usepackage{fancyhdr}
\usepackage{mathrsfs}
\usepackage{amssymb}
\usepackage[toc,page]{appendix}
\usepackage{xcolor}
\usepackage{float}
\usepackage{lmodern}
\usepackage{amsmath}
\usepackage{tensor}
\usepackage{lmodern}
\usepackage{amsmath}
\usepackage{tensor}
\usepackage{cite}

\begin{document}

\title{A class of static spherically symmetric solutions in $f(Q)$-gravity}

\author{ 
Marco Calz\'a$^1$\footnote{E-mail address: mc@student.uc.pt
},\,\,\,
Lorenzo Sebastiani $^{2}$\footnote{E-mail address:lorenzo.sebastiani@unitn.it}
\\
\\
\begin{small}
$^1$ CFisUC, Departamento de Fisica, Universidade de Coimbra, 3004-516 Coimbra,
Portugal
\end{small}\\
\begin{small}
$^2$ Dipartimento di Fisica, Universit\`a di Trento, Via Sommarive 14, 38123 Povo (TN), Italy
\end{small}\\
}

\date{}

\maketitle

\abstract{We analyze a class of topological static spherically symmetric vacuum solutions in 
$f(Q)$-gravity. We considered an Ansatz ensuring that those solutions trivially satisfy the field equations of the theory when the non-metricity scalar is constant. 
In the specific, we provide and discuss local solutions in the form of black holes and traversable wormholes. }

\section{Introduction}

The Equivalence Principle forces gravity to possess a geometrical nature. General Relativity (GR) is a geometric theory of gravity in which space-time is described as a Riemannian manifold, namely, a manifold in which the affine connection is the metric-compatible and torsion-free Levi-Civita connection totally defined by the metric. Thus, in GR, the scalar curvature $R$ is the fundamental quantity describing the manifold. However, the Riemannian geometry is an arbitrary request, and in general, a manifold is characterized by three fundamental geometrical objects: the curvature $R$, the torsion $T$, and the non-metricity $Q$  \cite{BeltranJimenez:2019esp}.
As a consequence, different theories of gravity can be built according to the properties of their connection. In particular, sub-classes of metric-affine geometry are dobbed as torsion-free ($T=0$), Riemann-Cartan ($Q=0$), and teleparallel ($R=0$). Moreover, further subsets are obtained if $Q$ and $T$, or $Q$ and $R$, or $R$ and $T$ vanish simultaneously. In those cases the geometries take the names of Riemannian ($T=Q=0$), Weitzenb\"ock or teleparallel ($R=Q=0$), and symmetric teleparallel ($R=T=0$). A last trivial subset is when the three quantities vanish together and the manifold is merely Minkowskian.

It is important to notice that there exist equivalent and alternative formulations of GR. One is the Teleparallel Equivalent of GR (TEGR) \cite{Springer,Maluf:2013gaa, moltoaltro}, equipped with a Weitzenb\"ock connection and vanishing curvature and non-metricity. A second equivalent formulation is
the Symmetric Teleparallel Equivalent of GR (STEGR)
\cite{Nester:1998mp,Adak:2004uh,Adak:2005cd,Adak:2008gd,Mol:2014ooa,BeltranJimenez:2017tkd,BeltranJimenez:2018vdo,Gakis:2019rdd} in which both curvature and torsion vanish.
Similarly to GR, in those equivalent theories, the Lagrangian density coincides with the respective scalars $T$ and $Q$.
Recent reviews and comparisons of these equivalent formulations can be found in Refs. \cite{Jarv:2018bgs,Capozziello,Heisenberg:2018vsk}.

Despite GR's successful results in describing the Solar System and larger structures of our Universe, many open problems including the dark contents of the Universe, the early-time inflation, and the difficulties in its quantization remain. Today, it is well accepted that GR (or its equivalent formulations) may be not the ultimate theory of gravity and some modifications may be required. Thus, when one deals with GR, TEGR, or STEGR, the simplest modification consists of changing the Lagrangian density allowing it to depend on a function of respective curvature, torsion, or non-metricity scalars. In this way one obtains the so-called theories of $f(R)$-gravity  \cite{DeFelice:2010aj,Sotiriou:2008rp}, $f(T)$-gravity \cite{Cai:2015emx,Bahamonde:2021gfp, fT1, fT12, fT13, fT41, fT42, SariRep, mioultimo} and $f(Q)$-gravity \cite{BeltranJimenez:2017tkd,BeltranJimenez:2018vdo,Zhao:2021zab,Lazkoz:2019sjl,Mandal:2020lyq, Mandal,Capozziello:2022tvv, Capozziello:2022wgl}.
Recently, the similarities to and differences from $f(R)$, $f(T)$, and $f(Q)$ gravity in terms of symmetry breaking and degrees of freedom have also been analyzed \cite{Hu:2022anq}.

Finding exact (vacuum) Static Spherically Symmetric (SSS) solutions in a gravitational theory is an interesting task, especially in light of the last observations. The
LIGO/Virgo Collaboration detection of gravitational waves from Black Hole (BH) coalescence \cite{LIGOScientific:2016aoc} and the Event Horizon Telescope direct observation of the BH shadows at the center of the Milky Way \cite{EventHorizonTelescope:2022xnr} and of M87 \cite{EventHorizonTelescope:2019dse}, represent new possibilities to test gravity in a strong-field regime \cite{Banerjee:2019nnj,Mark:2017dnq,Bueno:2017hyj}.
Those observations motivate the research of local solutions in the form of BHs, Wormholes (WHs), or ultra-compact stars, and the simplest description of these objects is provided by SSS geometries. In this respect, many local solutions have been already explored in $f(R)$-gravity \cite{Multamaki:2006zb,delaCruz-Dombriz:2009pzc,Kobayashi:2008tq,Upadhye:2009kt,Calza:2018ohl}, in $f(T)$-gravity \cite{Aftergood:2014wla,Bahamonde:2019jkf,Ilijic:2018ulf,LinfT,DeBenedictis:2016aze} and lately also in $f(Q)$-gravity \cite{Lin,Hassan:2022jgn,Tayde:2022lxd,Sokoliuk:2022efj,Banerjee:2021mqk,Hassan:2022hcb,Maurya:2022wwa,Parsaei:2022wnu,Wang:2021zaz,Mandal:2021qhx,DAmbrosio:2021zpm,Bahamonde:2022esv,Bahamonde:2022zgj}.

In Ref. \cite{Calza:2018ohl} SSS space-time vacuum solutions in the framework of $f(R)$-gravity has been explored for a special class of models where (on shell) the Lagrangian density vanishes and the Ricci scalar $R$ is a constant. Here, in an analog way, we consider (topological) SSS vacuum solutions in 
$f(Q)$-gravity models where the Lagrangian density vanishes and the non-metricity scalar $Q$ is a constant.  
With these requirements, the equations of motion are automatically satisfied in vacuum without solving them explicitly. As a consequence, a wide class of new solutions can be found and we discuss several possibilities by focusing our attention on BH and WH description.

The paper is organized as follows. In Sec. {\bf 2} we revisit the formalism of $f(Q)$-gravity. In Sec. {\bf 3} we introduce a class of $f(Q)$-models with exact SSS solutions for constant non-metricity scalar and we provide
some examples of Lagrangians. Sec. {\bf 4} and Sec. {\bf 5}
are devoted to the study of vacuum topological BH and WH solutions, respectively. 
Conclusions and final remarks are given in
Sec. {\bf 6}.

We use units of $k_B = c = \hbar = 1$ and we denote the gravitational constant $\kappa^2 = 8\pi G_N$.

\section{$f(Q)$-gravity}

In this section, we outline some general features of $f(Q)$-gravity. We will restrict our description in terms of components (the reader will find a more rigorous derivation in terms of forms in Ref. \cite{Lin}). 

The general affine connection of a parallelizable and differentiable manifold reads,
\begin{equation}\label{affine}
    \Gamma^\sigma_{\;\mu \nu}= \tilde \Gamma^\sigma_{\;\mu \nu} + K^\sigma_{\;\mu \nu} + L^\sigma_{\;\mu \nu}\,.
\end{equation}
Here, $\tilde \Gamma^\sigma_{\;\mu \nu}$  is the well-known Levi-Civita connection defined by the metric as
\begin{equation}\label{Levi-Civita}
  \tilde \Gamma^\sigma_{\;\mu \nu} = \frac{1}{2} g^{\sigma \rho} \left( \partial_\mu g_{\rho \nu} + \partial_\nu g_{\rho \mu}- \partial_\rho g_{\mu \nu}\right)\;\;\;\;.
\end{equation}
Moreover, $K^\sigma\,_{\mu \nu}$ is the contortion 
\begin{equation}\label{contortion}
   K^\sigma_{\;\mu \nu}= \frac{1}{2}  T^\sigma_{\;\mu \nu}+ T^{\;\;\;\sigma}_{(\mu \;\; \nu)}\;\;\;\;,
\end{equation}
with $T^\sigma_{\; \mu \nu}=2 \Gamma^\sigma_{\; [ \mu \nu]}$ torsion tensor. Finally, $ L^\sigma_{\;\mu \nu}$ is the deformation and reads,
\begin{equation}\label{deformation}
   L^\sigma_{\;\mu \nu}= \frac{1}{2} Q^\sigma_{\;\mu \nu} - Q^{\;\;\;\sigma}_{(\mu \;\; \nu)}\;\;\;\;,
\end{equation}
where $ Q^\sigma_{\;\mu \nu}$ is the non-metricity tensor
given by
\begin{equation}\label{nonmetricity}
   Q_{\sigma \mu \nu}= \nabla_\sigma g_{\mu \nu}= \partial_\sigma g_{\mu \nu} -\Gamma^\rho_{\;\sigma \mu } g_{\nu \rho} - \Gamma^\rho_{\;\sigma \nu } g_{\mu \rho } \;\;\;\;.
\end{equation}
Therefore, the non-metricity scalar is,
\begin{align}\label{non-m scalar}
   Q=&g^{\mu, \nu}(L^\alpha_{\beta \nu}L^\beta_{\mu \alpha}-L^\beta_{\alpha \beta}L)^\alpha_{\mu \nu} \nonumber\\
     &Q_{\sigma \mu \nu} P^{\sigma \mu \nu} \;\;\;\;,
\end{align}
where  $P^{\sigma \mu \nu}$ is the non-metricity conjugate given by,
\begin{equation}\label{non-m conjugate}
   P^\sigma_{\; \mu \nu} = \frac{1}{4} \left(-Q^\sigma_{\;\mu \nu}+2Q^{\;\;\;\sigma}_{(\mu \;\; \nu)} + Q^\sigma g_{\mu \nu}- \tilde{Q}^\sigma g_{\mu \nu} - \delta^\sigma_{(\mu} Q_{\nu)}\right)\;\;\;\;,
\end{equation}
with $Q_\sigma=Q^{\;\;\mu}_{\sigma \;\; \mu}$, and $\tilde{Q}_\sigma=Q^{\mu}_{ \;\; \sigma \mu}$.

If torsion and non-metricity are null, the connection is equivalent to the Levi-Civita connection which is metric-compatible. In symmetric teleparallel gravity, curvature and torsion are null and the non-metricity depends on metric and connection. 

Modified symmetric teleparallel gravity was introduced in Ref. \cite{BeltranJimenez:2017tkd} and the action reads,
\begin{equation}\label{action}
    I=-\frac{1}{2\kappa^2}\int_\mathcal{M}   f(Q) \sqrt{-g} d^4x + \int_\mathcal{M}  \mathcal{L}_m  \sqrt{-g}  d^4x \,,
\end{equation}
where
$g$ is the determinant of the metric tensor $g_{\mu\nu}$, $\mathcal M$ is the space-time manifold,
$f(Q)$ is a generic function of the non-metricity scalar $Q$ and
$\mathcal{L}_m$ is the Lagrangian density of matter contents.

In order to obtain the field equations of the theory one applies to (\ref{action}) independent variations with respect to both the metric and the connection, having so
\begin{equation}\label{1st EOM}
   \frac{2}{\sqrt{-g}} \nabla_\alpha \left( \sqrt{-g} f_Q P^\alpha_{\;\; \mu \nu }\right) + \frac{1}{2} g_{\mu \nu} f + f_Q \left( P_{\mu \alpha \beta } Q^{\;\;\alpha \beta}_\nu - 2P_{\alpha \beta \mu} Q^{\;\;\alpha \beta}_\nu\right)= \kappa^2 \mathcal{T}_{\mu \nu}\;\;\;\;,
\end{equation}
\begin{equation}\label{2nd EOM}
    \nabla_\mu \nabla_\nu \left(\sqrt{-g} f_Q P^\alpha_{\;\; \mu \nu }\right)=0\;\;\;\;,
\end{equation}
where $\mathcal{T}_{\mu \nu}$ is, as usually, the energy-momentum tensor of matter, namely
\begin{equation}\label{EMT}
   \mathcal{T}_{\mu \nu}= -\frac{2}{\sqrt{-g}} \frac{\delta(\sqrt{-g}\mathcal{L}_m)}{\delta g^{\mu \nu}}\;\;\;\;.
\end{equation}
In the expression above, $f\equiv f(Q)$ and $f_Q=\frac{d f(Q)}{dQ}$.
We note that the Lagrangian density of matter is taken independently with respect to the connection, so no hyper-momentum appears. Moreover, at is well known, one obtains the results of GR (in STEGR framework) by posing $f(Q)=Q$, such that the Lagrangian density reads $\mathcal L=-\frac{Q}{2\kappa^2}+\mathcal L_m$.

\section{The model Ansatz}

Let us consider topological SSS space-time in the form,
\begin{equation}
ds^2=-h(r)dt^2+\frac{1}{g(r)} dr^2+r^2\,\left(\frac{d\rho^2}{1-k\rho^2}+\rho^2 d\phi^2\right) \,,\label{metric}
\end{equation}
where $h(r)$ and $g(r)$ are functions of the radial coordinate $r$ only
and the two-dimensional space $d\Omega_k^2=\frac{d\rho^2}{1-k\rho^2}+\rho^2 d\phi^2 $ may assume three different topologies, depending on the choice $k=1,0,-1$. Namely, 
    the manifold will be either a sphere $S_2$, a torus $T_2$ or a compact hyperbolic manifold $Y_2$, according to whether $k = 1,0,-1$, respectively.
   Metrics of this type can be denoted as Static Pseudo-Spherically Symmetric space-time,
but often are simply called SSS space-time.

In this paper we use the prescription introduced in Ref. \cite{Lin} where a detailed discussion about the covariant formulation of the theory can be found. Since the vanishing of curvature and torsion forces the connection $\Gamma^\sigma_{\mu\nu}$ to be purely inertial, its only non-vanishing components correspond to the non-vanishing components of Levi-Civita connection in the absence of gravity and for metric (\ref{metric}) we get
\begin{align}
&\Gamma^r_{\phi\phi}=- k r
\rho^2,  \;\;\;\;\; \Gamma^\rho_{r\rho}=\Gamma^\phi_{r\phi}=\frac{1}{r},  \;\;\;\;\;\Gamma^\phi_{\rho\phi}=\frac{1}{\rho},\nonumber\\
&\Gamma^r_{\rho\rho}= -\frac{k r}{1-k \rho^2},  \;\;\;\;\;\Gamma^\rho_{\rho\rho}= \frac{k \rho}{1-k \rho^2},  \;\;\;\;\;\Gamma^\rho_{\phi\phi}=-\rho(1-k \rho^2)
\,.
\end{align}\\
Thus, the non-vanishing $Q_{\alpha\beta\gamma}$ and $L^{\alpha}_{\beta\gamma}$ are given by,
\begin{align}
&Q_{r t t}=-h'(r),\;\;\;\;\;Q_{r r r}=-\frac{g'(r)}{g^2(r)}, \nonumber\\
&Q_{\rho r \rho}=Q_{\rho \rho r}=-\frac{r-\frac{k r}{g(r)}}{(1- \rho^2)}, \nonumber\\
&Q_{\phi r \phi}=Q_{\phi \phi r}= \left(\frac{1}{g(r)}-1\right) r \rho^2, 
\end{align}
\begin{align}
&L^t_{t r}=L^t_{r t}=-\frac{h'(r)}{2h(r)}\;\;\;L^r_{t t}=-\frac{1}{2}g(r)h'(r),\;\;\;\;\;L^r_{r r}=\frac{g'(r)}{2g(r)}, \nonumber\\
&L^r_{\rho \rho}=-\frac{(k-g(r))r}{(1-k \rho^2)}, \;\;\;\;\; L^r_{\phi \phi}=(g(r)-k)r \rho^2\,. 
\end{align}
Finally, the non-metricity scalar can be computed as,
\begin{equation}
Q=\frac{\left(g(r)-k\right)\left(\frac{h'(r)}{h(r)}-\frac{g'(r)}{g(r)}\right)}{r }\,.
\label{Q}
\end{equation}
In Ref. \cite{Lin} SSS solutions in the framework of $f(Q)$-gravity has been already investigated in both, vacuum and non-vacuum cases. In particular, it has been found that, given a generic model of $f(Q)$-gravity, the only solution in vacuum is the Schwarzschild de Sitter/Anti-de Sitter (dS/AdS) one. Here, we point out that, for a special class of $f(Q)$-gravity models, new vacuum solutions can be realized.

In fact, if $T_{\mu\nu}=0$, Eqs. (\ref{1st EOM})--(\ref{2nd EOM}) are automatically satisfied under the assumption
\begin{equation}
f(Q_0)=f_Q(Q_0)=0\,,
\label{Ansatz}
\end{equation}
where $Q_0$ is a constant value  (eventually null) of the non-metricity scalar and is related to 
the metric functions $h(r)\,,g(r)$ through Eq. (\ref{Q}), namely
\begin{equation}
\frac{\left(g(r)-k\right)\left(\frac{h'(r)}{h(r)}-\frac{g'(r)}{g(r)}\right)}{r }=Q_0\,.
\label{Q0}
\end{equation}
Despite to the restriction to fulfill the Ansatz (\ref{Ansatz}) for some value of non-metricity $Q_0$, the functional form of $f(Q)$ is not uniquely determined and among all the possible choices we may find some cases of modified symmetric teleparallel gravity of physical interest.

In analogy with $f(R)$-gravity, higher-order corrections incur deviations from Standard Model in the past\footnote{On Friedmann-Robertson-Walker space-time we have $Q=6H^2$, with $H$ Hubble parameter.}
playing a role in the description of the early-time Universe and supporting inflation or bounce cosmology.
In this respect, polynomial models in the form (here, $\gamma$ is a dimensional constant),
\begin{equation}
f(Q)=\gamma (Q-Q_0)^n\,,\quad 2\leq n\,,    
\end{equation}
satisfy condition (\ref{Ansatz}). We can expand the polynomial as,
\begin{eqnarray}
f(Q)&=&\gamma \sum_{k=0}^{n}
\frac{n!}{k!(n-k)!}Q^{n-k}(-Q_0)^k\,,
\nonumber \\ 
&=&\gamma \frac{n!}{(n-1)!}(-Q_0)^{n-1} Q
+\gamma (-Q_0)^n+
\gamma
\sum_{k=0}^{n-2}
\frac{n!}{k!(n-k)!}Q^{n-k}(-Q_0)^k\,.
\label{mod0}
\end{eqnarray}
If $\gamma\frac{n!}{(n-1)!}(-Q_0)^{n-1}=1$, the linear term corresponds to STEGR formulation of GR, and if we assume $\gamma\ll 1$ the powers of $Q$ are relevant only for large values of it when $Q\sim Q_0$. Note that also a cosmological constant term appears, but some suppressing mechanism can be easily introduced without destroying the feature of the model. For example,
by starting from the second row of (\ref{Ansatz}),
we can rewrite the Lagrangian by replacing $\gamma\rightarrow \gamma(Q)$, where $\gamma(Q)$ is a function of non-metricity scalar, in front of the cosmological constant term $(-Q_0)^n$ only. If $\gamma(Q_0)=1$ and $\gamma'(Q_0)=0$, the model still satisfies (\ref{Ansatz}). The function $\gamma(Q)$ can now be constructed in order to vanish at low red-shift when $Q\ll Q_0$. For instance, we may take
\begin{equation}
\gamma(Q)=   
\text{e}^{-\beta(Q-Q_0)^{2m}}\,,\quad
\beta>0\,,m>1\,.
\end{equation}
Other examples of Lagrangians that satisfy condition (\ref{Ansatz}) for a given value of $Q_0$ and whose implementation makes sense at large values of non-metricity scalar can be constructed with trigonometric and hyperbolic functions as:
\begin{eqnarray}
f(Q)&=&\gamma \left(\cos(\beta(Q-Q_0))-1 \right)\,,\nonumber\\
f(Q)&=&\gamma \left(\cosh(\beta(Q-Q_0))-1 \right)\,.
\end{eqnarray}
Here, $\gamma$ and $\beta$ are dimensional constants. In this case, the STEGR formulation of GR plus higher order corrections is recovered after Taylor expansions around $Q\simeq Q_0$.

In the context of dark energy phenomenology, we can consider
the following
Lagrangian,
\begin{equation}
f(Q)=Q+2\Lambda\left(1-\text{e}^{\frac{Q}{2\Lambda}}\right)\,,    
\end{equation}
where $\Lambda$ is the cosmological constant. This model falls in the class of theories under investigation, since the condition (\ref{Ansatz}) is satisfied when $Q_0=0$ and offers the equivalent description of one-step models of $f(R)$-gravity \cite{onestep, onestep2,onestep3, onestep4,onestep5,onestep6,onestep7,onestep8,onestep9,onestep10}, since for large values
of $Q$ we recover the results of STEGR plus cosmological constant with $f(Q)\simeq Q+2\Lambda$.

In the following sections, we will discuss BH and WH solutions which always are admitted as vacuum solutions in these models of $f(Q)$-gravity.   The required and common propriety of the various metrics is that they lead to a constant value of non-metricity scalar which is fixed by the gravitational Lagrangian itself and which allows to trivially solve the field equations of the theory.

\section{Black hole solutions}

In this section, we are interested in topological black hole solutions which satisfy Eq. (\ref{Q0}) for some value of $Q_0$. Thus, it is useful to replace $h(r)=\text{e}^{2\alpha(r)}g(r)$ inside the metric (\ref{metric}), in order to obtain,
\begin{equation}
ds^2=-\text{e}^{2 \alpha(r)} g(r)dt^2+\frac{1}{g(r)} dr^2+r^2\,d\Omega_k^2 \,,\label{metricBH}
\end{equation}
Here, $\alpha(r)$ and $g(r)$ are functions of the radial coordinate $r$ only. In general, a zero of $g(r)$, namely a value of $r=r_0$ for which $g(r_0)=0$,  defines an event horizon as soon as $g'(r_0)>0$, such that one deals with a positive surface gravity.

At first, we observe that when  $\alpha(r)=0$ (the result can be easily generalized to the case $\alpha=const$ with a constant rescaling of time coordinate), the non-metricity scalar $Q$ is identically null. Namely, Eq. (\ref{Q0}) holds true for $Q_0=0$. This is fully consistent with Ref. \cite{Lin}, where it is stressed that the Schwarzshild solution makes the non-metricity scalar vanish and satisfies the vacuum field equations in the framework of $f(Q)$-gravity. Here, we find a more general result for our class of Lagrangians. Given a topological SSS solution in the form
\begin{equation}
  ds^2=- g(r)dt^2+\frac{1}{g(r)} dr^2+r^2\,\left(\frac{d\rho^2}{1-k\rho^2}+\rho^2 d\phi^2\right) \,,\label{metricBH2} 
\end{equation}
the special class of models with the propriety
\begin{equation}
f(0)=0\,,\quad f_Q(0)=0\,,    
\end{equation}
admits this space-time as vacuum solution for any choice of $g(r)$. For example, in the quadratic model $f(Q)=Q^2$ any topological BH solution (\ref{metricBH2}) can be realized and some interesting scenarios may take place. Apart from the Schwarzshild BH solution, regular BHs 
free of central singularity can be reproduced without invoking exotic matter and avoiding in this way the instability related to the negative speed of sound which often appears in the framework of GR. Moreover, one may look for solutions describing rotation curves of galaxies, like, for example, the (topological) Riegert-inspired solution
\cite{Riegert} with
\begin{equation}
g(r)=    k+\frac{m}{r}+c_0 r\,,
\end{equation}
where $m$ is a mass parameter and $c_0$ is a constant. In this expression, we can read the contribution of baryonic matter content and phenomenological dark matter content, separately. The first one corresponds to the classical Newtonian potential, while the second one is linear with respect to the radial coordinate and can reproduce the observed flattening of the rotation curves \cite{DM1, DM2}.

On the other side, when $\alpha(r)$ in the metric (\ref{metricBH}) is not a constant, Eq. (\ref{Q0}) leads to
\begin{equation}
g(r)=\frac{2 k\alpha'(r) + Q_0 r}{2 \alpha'(r)}\,.  
\label{ciaone}
\end{equation}
Here, we note that only in the spherical topological case with $k=1$, SSS solutions having $Q_0=0$ when $\alpha(r)\neq \text{const}$ can be realized. In this case
$g(r)=1$, $\alpha(r)$ is an arbitrary function, and the solution can not describe a BH. 
Namely, BH space-times
with zero non-metricity scalar 
are possible only when $\alpha(r)=\text{const}$ and therefore $g_{00}=g_{11}^{-1}$.  

On the other side, when $Q_0\neq 0$, various examples of ``dirty'' black holes can be found by considering
\begin{equation}
\alpha(r)=-\frac{Q_0 r^{2+z}}{2m(2+z)}\,,    
\end{equation}
such that
\begin{equation}
g(r)=k-\frac{m}{ r^z}\,,    
\end{equation}
with $m$ a constant and $z\neq -2$ a real number. In order to obtain $g_{00}\equiv - \text{e}^{2\alpha(r)} g(r)>0$ for $r\rightarrow 0^+$, we may require $m\,,z>0$ when $k=1$ and 
$z<0$ when $k=-1$.
Thus, for the spherical topological case with $k=1$, we obtain a BH where the event horizon is located at $r_H=(m)^\frac{1}{z}$ such that $g'(r_H)>0$. The same happen for the hyperbolic case with $k=-1$ when $m<0$, where the BH horizon is located at  $r_H=(-m)^\frac{1}{z}$ with $g'(r_H)>0$.

One recovers a topological Reissner-Nordstr\"om-like solution by taking
\begin{equation}
\alpha(r)= 
-Q_0\left(\frac{r^3}{6m}
+\frac{c_0r^2}{4m^2}+\frac{c_0^2 r}{2m^3}+
\frac{c_0^3}{2m^4}\log(c_0-m r)
\right)\,,\label{ciao}
\end{equation}
which leads to
\begin{equation}
g(r)=k-\frac{m}{r}+\frac{c_0}{r^2}\,,    
\end{equation}
where $m$ is a mass parameter and $c_0$ and $c_1$ are constants. In general, the metric may
describe a black hole with the appearance of two horizons, an external event horizon at $r=r_+$ and an internal Cauchy horizon at $r=r_-$. For example, by taking $k=1$ and $m\,,c_0>0$, by solving the equation $g(r_\pm)=0$, one has
\begin{equation}
r_\pm=\frac{1}{2}
\left(
m\pm\sqrt{m^2-4c_0}
\right)\,,
\end{equation}
with $g'(r_+)>0$ and $g'(r_-)<0$.
Now, let us have a look at the role of the metric function $\alpha(r)$ in (\ref{ciao}).
We assume $Q_0<0$. The metric function $h(r)$ in (\ref{metric}) reads,
\begin{equation}
h(r)=\text{e}^{2\alpha(r)}g(r)=
\text{e}^{-Q_0\left(\frac{r^3}{6m}
+\frac{c_0r^2}{4m^2}+\frac{c_0^2 r}{2 m^3}
\right)}
(c_0-mr)^{\frac{-Q_0 c_0^3}{2 m^4}} g(r)\,,
\end{equation}
where we have used (\ref{metricBH}) and  (\ref{ciao}). 
The range of $r$ is restricted by the condition
$-g>0$ (we remember that $g$ is the determinant of the metric), namely
$\frac{h(r)}{g(r)}$ must be positive and real. Thus, 
if $\frac{c_0}{m}>0$, we have that
$r_0=\frac{c_0}{m}$ represents a sort of cut-off of the space-time and $r>r_0$. In particular, when $\frac{c_0}{m}<r_+$, we are still dealing with a BH solution having an external event horizon and in some cases an internal Chaucy horizon, but the central singularity at $r=0$ is avoidable. A similar mechanism is at the basis of the so-called black-bounce space-time
\cite{BB, BB2, BB3, BB4} and represents an interesting way to make the metric regular without the introduction of a de Sitter core (see also Ref. \cite{Zerg}). 

Finally, we can consider the  Lifshitz-like solutions with
\begin{equation}
\alpha(r)=\frac{1}{2}\log\left(\frac{r}{r_0}\right)^z\,,    
\end{equation}
where $r_0$ is a lenght-scale and $z\neq 0$ is a real number. From Eq. (\ref{ciaone}) we find
\begin{equation}
g(r)=k+\frac{Q_0 r^2}{z}\,,
\end{equation}
which is nothing else than a pure dS/Anti dS solution, depending on the sign of $Q_0$.

\section{Wormhole solutions}

We will now analyze some wormhole solutions \cite{Morris} which satisfy the condition (\ref{Q0}) for some value of $Q_0$, namely which can be realized as vacuum solution of $f(Q)$-models under the Ansatz (\ref{Ansatz}). We remember that in GR, traversable wormholes take place only in the presence of a matter source violating the null energy condition \cite{WH,WH2,WH3,WH4,WH5,WH6,WH7,WH8,WH9,WH10,WH11}, while here the role of the anti-gravitational matter can be played by the modification of gravity itself (see also Refs. \cite{WHMG, WHMG2, WHMG3,WHMG4}).

For our purpose, it is convenient to rewrite the metric function $h(r)$ in (\ref{metric}) as $h(r)=\text{e}^{2\Phi(r)}$, such that 
\begin{equation}
ds^2=
-\text{e}^{2\Phi(r)} dt^2+\frac{1}{g(r)} dr^2+r^2\,\left(\frac{d\rho^2}{1-k\rho^2}+\rho^2 d\phi^2\right) \,,\label{metricWH}
\end{equation}
where $\Phi(r)$ is dubbed red-shift function and, as well as $g(r)$, it depends on the radial coordinate $r$ only.

A traversable wormhole occurs if the radial coordinate is embedded by a minimal radius or ``throat'' at $r=r_0$, in order to prevent the appearance of horizons. The function $\Phi(r)$ needs to be finite and regular everywhere along the throat and the following conditions must be satisfied \cite{WHcond,WHcond2,WHcond3}:
\begin{itemize}
\item $\Phi_\pm(r)$ and $g_\pm(r)$ are well defined for all $r \geq r_0$;
\item  $\Phi'_+(r_0)=\Phi'_-(r_0)$; 
\item $g_\pm(r_0)=0\,, g_\pm(r)>0$ for all $ r \geq r_0$;
\item $g'_+(r_0)=g'_-(r_0)>0$.
\end{itemize}
In general, the coordinate $r$ is ill-behaved near the throat,
but the proper radial distance is well defined everywhere as \cite{Morris}
\begin{equation}
l(r)=\pm\int ^r_{r_0}\frac{d\tilde r}{g(\tilde r)} \,.
\end{equation}
The minimal value of the proper distance is reached for $r=r_0$, while the positive and negative values of $l\equiv l(r)$ correspond to the lower and upper universes connected through the throat of the wormhole. Thus, the traveling time necessary to cross the wormhole between $l(r_1)=-l_1<0$ and $l(r_2)=+l_2>0$ results to be
\begin{equation}
\Delta t=\int_{l_1}^{l_2}   \frac{d l}{v \text{e}^{\Phi(l)}} \,.
\end{equation}
Here, $v=d l/[\text{e}^{\Phi(l)} dt]$ is the radial velocity of the traveler when he/she passes a given radius $r$.  The magnitude of $\Phi'(r)$ is associated with the tidal force experimented crossing the throat and big values of its modulus may make it difficult for an observer to complete the journey from one side to the other of the wormhole.

Firstly, we consider vacuum WH solutions with
$\Phi(r)$ constant in (\ref{metricWH}),
\begin{equation}
\Phi(r)=\text{const}\,.    \label{WH0}
\end{equation}
This choice corresponds to a vanishing tidal force such that the proper time measured by a static observer coincides with the time coordinate $t$. In this case Eq. (\ref{Q0}) leads to the following solutions for the metric function $g(r)$ when $ k =\pm 1 $,
\begin{equation}
g(r)=-k\, \text{W}\left[\frac{-\text{e}^{-\frac{c_0 }{k}+\frac{ Q_0 r^2}{2k}}}{k}\right]\,,\quad\text{when}\,\, Q_0\neq0\,, \nonumber
\end{equation}
\begin{equation}
g(r)=k\,,\quad \text{or}\quad \,g(r)=c_1\,,\quad\text{when}\,\, Q_0=0\,,
\end{equation}
where $c_0\,,c_1$ are generic constants.
In the first expression, $W\equiv W[x]$ is the principal solution of the Lambert function. When we restrict to real values the Lambert function becomes the inverse function of $x=W \text{e}^W $. The solutions for $Q_0=0$ are acceptable only in the spherical case  $k=1$ or for $c_1>0$, and they turn out to be Minkowski space-times (in the second case the areal radius is rescaled as $\mathcal R(r)= c_1 r$).  

More interesting is the flat case with $k=0$, for which we get
\begin{equation}
g(r)= c_0-\frac{Q_0 r^2}{2}\,,   \label{WH00}
\end{equation}
with $c_0$ constant. The solution describes a WH as soon as $c_0\,, Q_0<0$. In this case, the radial coordinate reaches a minimal value on the throat of the WH at $r_0=\sqrt{\frac{2c_0}{Q_0}}$
with $g'(r_0)>0$. 

This result can be generalized to the case where the metric function $\Phi(r)$ in (\ref{metricWH}) assumes the form,
\begin{equation}
\Phi(r)=\frac{1}{2}\log\left(\frac{r}{r_0}\right)^z\,,    
\end{equation}
with $r_0$ a length scale and $z\neq 0$ a constant parameter. Then, for $k=\pm 1$, if $Q_0=0$ Eq. (\ref{Q0}) can be solved  as
\begin{equation}
  g(r)=k\,,\quad \text{or}\,\,g(r)=c_1 r^z\,. 
\end{equation}
The corresponding solutions can be easily written in Gaussin-polar coordinates as,
\begin{equation}
ds^2=- h(r)dt^2+dr^2+\mathcal R^2(r)  
\,\left(\frac{d\rho^2}{1-k\rho^2}+\rho^2 d\phi^2\right)\,.
\end{equation}
In the first case, when $g(r)=k$, only the spherical topology with $k=1$ is admitted and $\mathcal R(r)=r$. 
In the second case, when $g(z)=c_1 r^z$,
by assuming $c_1>0$ and $z<2$, after 
the radial coordinate transformation $r\rightarrow \left(\frac{\sqrt{c_1}(2-z)r}{2}\right)^{\frac{2}{(2-z)}}$,
we obtain
\begin{equation}
ds^2=-\left(\frac{\sqrt{c_1}(2-z)r}{2r_0^{\frac{(2-z)}{2}}}\right)^\frac{2z}{(2-z)}dt^2
+dr^2+\mathcal R^2(r)
\,\left(\frac{d\rho^2}{1-k\rho^2}+\rho^2 d\phi^2\right)\,,\quad \mathcal R(r)=
\left(\frac{\sqrt{c_1}(2-z)r}{2}\right)^\frac{2}{2-z}\,.
\end{equation}
Moreover, when $z= 2$ and $c_1>0$,
by posing $r\rightarrow \text{e}^{\sqrt{c_1}r}$,
we have
\begin{equation}
   ds^2=-\frac{\text{e}^{2\sqrt{c_1}  r}}{r_0^2}dt^2+dr^2+ \mathcal R^2(r)
\,\left(\frac{d\rho^2}{1-k\rho^2}+\rho^2 d\phi^2\right)\,,\quad \mathcal R(r)=\text{e}^{\sqrt{c_1}r}\,.
\end{equation}
The metrics above clearly do not describe a WH. Note that when $0<z\leq 2$ the metric is regular everywhere, and the singularity at $r=0$ is removed with the change of coordinates, otherwise, when $z<0$, the metric function $h(r)$ diverges at $r=0$.

As in the previous case, interesting solutions can be found for the flat topological case with $k=0$, for which Eq. (\ref{Q0}) leads to,
\begin{equation}
g(r)=c_0 r^z-\frac{Q_0 r^2}{(2-z)}\,,    
\end{equation}
where $c_0$ is an integration constant.
In this case, WH solutions can be realized for negative values of $Q_0$ when $z<2$ and $c_0<0$ or when $z>2$ and $c_0>0$. In both cases the throat is located at $r_0=\left(\frac{Q_0}{(2-z)c_0}\right)^{\frac{1}{z-2}}$ such that $g'(r_0)>0$. In the limit $z=0$ we recover the solution given by (\ref{WH0}) and (\ref{WH00}).

Thus, in the special class of models satisfying the Ansatz (\ref{Ansatz}) for some negative value of $Q_0$, WHs with flat topology can be realized as vacuum solutions of the theory.

\section{Conclusions}

In this paper, we studied topological Static Spherically Symmetric vacuum solutions in a special class of $f(Q)$-gravity models. $f(Q)$-gravity is a modified theory of symmetric teleparallel gravity, where the Lagrangian density is a function of the non-metricity scalar $Q$ only. When the Lagrangian is proportional to $Q$, we recover the symmetric teleparallel formulation of GR.  Finding exact vacuum solutions in a gravity theory is an interesting task, especially in the presence of high-order field equations, as it occurs for $f(Q)$-gravity. We investigated a class of metrics leading to a constant non-metricity scalar value $Q_0$. Thus, all the models satisfying (on-shell) the simple Ansatz,
$f(Q_0)=f'(Q_0)=0$, admit these metrics as solutions. We stress that the functional form of $f(Q)$ is not uniquely determined and many Lagrangians of physical interest belong to this class of models. We also remark that in some previous papers (for example in Ref. \cite{Lin}) SSS solutions in general models of $f(Q)$-gravity have been investigated, but taking into account our Ansatz, we are able to find new solutions which suitable describe compact objects such as black holes and wormholes.

In the case of BH solutions, in addition to the Schwarzschild space-time, we can generate a large class of regular BH solutions, where the central singularity is absent. Furthermore, the simple metric requirement $g_{00}=g_{11}^{-1}$ is enough to ensure a zero non-metricity scalar and allows the construction of SSS solutions with any desired features. For example, we were able to identify solutions describing the anomalous rotation of spiral galaxies. Moreover, ``dirty'' BHs where $g_{00}\neq g_{11}^{-1}$ are also investigated. 

Traversable wormholes are hypothetical objects connecting two spacetime regions and guaranteeing that an observer may traverse them in a finite time. The usual theoretical downside of such solutions dwells in their demand for a source violating the null energy conditions.
With our analysis, we are able to identify a few traversable configurations characterized by the conceptual advantage that those solutions are in vacuum and do not require the presence of any matter sources violating the null energy condition. In particular, we found some interesting configurations for the flat topological case.

Nevertheless, at this stage, for both such BH and WH solutions, nothing can be said about the precise form of the function $f(Q)$ of the Lagrangian besides the requirement that it has to satisfy the constraint $f(Q_0)=f'(Q_0)=0$.
This condition selects solutions with constant non-metricity scalar and shows that the solution space is ample and deserves to be studied since provides a rich phenomenology.

\section*{Acknowledgements}
This work was supported by national funds from FCT, I.P., within the projects UIDB/04564/2020, UIDP/04564/2020 and the FCT-CERN project CERN/FIS-PAR/0027/2021. \newline
M.C. is also supported by the FCT doctoral grant SFRH/BD/146700/2019.

\end{document}